\documentclass[11pt]{article}
\usepackage{amsfonts,amssymb,amsmath,latexsym,eucal,epsfig,color,graphicx,dcolumn,subfigure}
\begin{document}

\begin{titlepage}
    \begin{flushright}
    {UTAP-571}
    \end{flushright}
    \vspace{2.5cm}
    \begin{center}
	{\Large{\bf 
        	Doubly Spinning Black Rings 	
	}      }
    \end{center}
    \vspace{1cm}
    \begin{center}
    Hideaki Kudoh
    \\
    \vspace{1.0cm}
    {\small \textit{
    Department of Physics, UCSB, Santa Barbara, CA 93106, USA
\\
    and
\\
Department of Physics, The University of Tokyo, Tokyo 113-0033, Japan
    }}
\\
\vspace*{1.0cm}
    {\small
    {\tt{
    kudoh at physics.ucsb.edu
    }}
    }
\end{center}
\vspace*{1.0cm}

\begin{abstract}
We study a method to solve stationary axisymmetric vacuum Einstein equations numerically. 
As an illustration, the five-dimensional doubly spinning black rings that have two independent angular momenta are formulated in a way suitable for fully nonlinear numerical method. 
Expanding for small second angular velocity, the formulation is solved perturbatively upto second order involving the backreaction from the second spin. 
The obtained solutions are regular without conical singularity, and the physical properties are discussed with the phase diagram of the reduced entropy vs the reduced angular momenta. 
Possible extensions of the present approach to constructing the higher dimensional version of black ring and the ring with the cosmological constant are also discussed.
\end{abstract}



\end{titlepage}

\section{Introduction}

At the level of classical theory of gravity, black holes are simple but general vacuum state of spacetime itself.  
This is a noteworthy consequence in the four-dimensional black hole physics by the uniqueness theorems. 
Nevertheless, corresponding to the fact that the number of degrees of freedom of gravity increases with spacetime dimensions, such simplicity of spacetime seems to hold only in four-dimensions, as was realized by the discovery of five-dimensional black rings (BRs) with horizon topology $S^1 \times S^2$ \cite{Emparan:2001wn}. 
The BRs are in good contrast with the higher-dimensional Kerr solution (so-called Myers-Perry black holes), and many aspects have been studied so far including extensions of the solution, partial stability, role in string theory, etc \cite{Emparan:2006mm}.

At present, one of important aspects concerning BRs is to find out new type of ``black ring" in higher dimensions. 
In more than five dimensions, we expect that not only black ``rings" but also new types of topologies might exist. 
Developing a feasible effective method to construct such new black holes is an interesting open question. This will be also related to finding conjectured black holes with less symmetry than any known solutions \cite{Reall:2002bh}. 
Search for new black holes is important to understand the phase space of higher dimensional vacuum spacetimes.

To tackle the issue, we need a new approach or tool. 
After the original investigation of BRs which was mainly based on guesswork (and knowledge of four-dimensional solution~\cite{Dowker:1993bt}), several systematic methods which can rederive the vacuum black ring solution have been developed \cite{Mishima:2005id,Iguchi:2006rd,Iguchi:2006tu,
Tomizawa:2006vp,Tomizawa:2005wv,Koikawa:2005ia,Pomeransky:2005sj,Harmark:2005vn}. An outcome is the discovery of vacuum BRs with rotation on the $S^2$ but without rotation along the $S^1$ circle \cite{Mishima:2005id}. 
(An independent work based on tractable coordinates is \cite{Figueras:2005zp}.)
The five-dimensional BRs have two independent rotational planes and the two rotations can generate two independent angular momenta. 
Thus we expect doubly spinning BRs that have the rotation on both $S^2$ and $S^1$.
However, the exact solution of doubly spinning BRs has not yet been discovered, and no concrete procedure to find it has been available. 
The black ring with two angular momenta is known for supersymmetric case, but the two angular momenta are not independent conserved charges\cite{Elvang:2004rt}.

The present paper aims at exploring a new approach to search for new type of black holes.  
A numerical approach has been employed and developed to discover static axisymmetric solutions over the past several years~\cite{Wiseman:2002zc,Kudoh:2004hs,Kudoh:2003ki,Kudoh:2003xz,Sorkin:2003ka}. 
The advantage of numerical approach is that it will be widely applicable once the basic scheme and techniques are established, and therefore it will be interesting to extend the previous approach applied for static axisymmetric spacetimes to stationary ones. 
To construct a solution numerically, it needs adequate understandings of expected solution.  
However, as we will see, BRs are rather complicated in numerics to hold a consistency over whole system, and boundary conditions for unknown solutions are not necessarily clear a priori.
In numerics, we need to fix all necessary information before starting the calculation. This is in contrast with an analytic approach, in which we can impose necessary conditions after we find a general solution, roughly speaking.

In this paper, we adopt the canonical form of metric for stationary axisymmetric spacetimes \cite{Harmark:2004rm,Harmark:2005vn} as suitable coordinates for numerics, and we will work on five dimensions. 
Although the so-called ring-coordinates \cite{Emparan:2006mm} is the most adopted coordinate system for BRs, it is not rectangular at the asymptotic infinity (boundary) and less useful. 
As a good example to test such approach, we focus on the issue of doubly spinning BRs. 
We formulate the doubly spinning BRs in a numerically manageable manner, and solve it in a perturbative way. 
Based on the solutions, some of physical properties of the doubly spinning BRs are discussed. 
This specific example helps us to clarify subtle issues related to numerics, and we discuss how we can deal with fully nonlinear stationary solutions in the present approach.
Based on these, we also discuss a possible extension to find higher dimensional BRs with new topologies and BRs in AdS.

The paper is organized as follows.
In the next section, we introduce the canonical form of metric and explain necessary ingredients.
In Sec.~\ref{sec:Formulation of doubly spinning black rings}, we formulate doubly spinning black rings in a generic way and provide the rod-structure which works as a boundary condition. 
In Sec.~\ref{sec:Perturbative scheme}, we solve the doubly spinning BRs in a perturbative way and discuss the physical properties in details. 
Several possible extension of the present approach are discussed in Sec.~\ref{sec:Extensions of the approach}.  
The final section is devoted to a summary.

\section{Stationary axisymmetric spacetimes}
\label{sec:Stationaly axisymmetric spacetimes}

\subsection{Canonical form of metric}
\label{sec:Canonical form of metric}

Following to the formulation in \cite{Harmark:2004rm} (and \cite{Emparan:2001wk} for static case), we introduce the canonical form of metric and collect necessary ingredients for later discussions.
Stationary axisymmetric $D$-dimensional spacetimes can be put in the following form,
\begin{eqnarray}
  ds^2  
    = \sum^{D-2}_{a,b=1}  g_{ab} dx^a dx^b + e^{2\nu} (dr^2 + dz^2). 
\label{eq:conformal metric anstaz1}
\end{eqnarray}
If this spacetimes has $D-2$ commuting Killing vectors, the metric is in a special class. 
For such cases, the canonical coordinates in which the coordinate $r$ satisfies 
\begin{eqnarray}
  r^2={\det(-g)},
\label{eq:r2=detg}
\end{eqnarray}
are available. 
This canonical form of metric simplifies the Einstein equations considerably. 
The vacuum Einstein's equations are
\begin{eqnarray}
2e^{2\nu} 
R_{ab} &=& 
- \nabla^2 g_{ab}
+ g^{dc} \left(  \partial_r g_{ad} \partial_r g_{bc} \right)
+ g^{dc} \left(  \partial_z g_{ad} \partial_z g_{bc} \right),
\label{eq:R_ab}
\\
R_{zz} &=& 
- \nabla^2 \nu  
- \frac{1}{4} g^{ab} g^{cd } \partial_z g_{ac} \partial_z g_{b d} ,
\label{eq:R_zz}
\\
R_{rz}  &= &
  \frac{1}{r} \partial_z \nu
- \frac{1}{4} g^{ab} g^{cd } \partial_z g_{ac} \partial_r g_{b d} , 
\label{eq:R_rz}
%
%
\\
 R_{rr} - R_{zz} &=& 
\frac{1}{r^2} + \frac{2}{r} \partial_r \nu 
+ \frac{1}{4} g^{ab} g^{cd } (\partial_z  g_{ac} \partial_z g_{b d} 
-   \partial_r  g_{ac} \partial_r g_{b d} ),
\label{eq:R_rrRzz}
\end{eqnarray}
where we have introduced the operator  
$
\nabla^2 = \partial_r^2 + \partial_z^2 + \frac{1}{r} \partial_r,
$
which works as the Laplace operator in three-dimensional cylindrical coordinates. 
The first two differential matrix equations with the constraint (\ref{eq:r2=detg}) are the basic field equations to be solved. 
The last two equations are the constraint equations. 
It can be shown that the constraint equations are satisfied as long as 
(\ref{eq:R_ab}) and (\ref{eq:R_zz}) are satisfied. 
If we find a solution of (\ref{eq:R_ab}), $\nu(r,z)$ will be immediately obtained from (\ref{eq:R_zz}), and thereby a complete solution of the Einstein equations will be obtained.

\begin{figure}[t]
\begin{center}
\includegraphics[width=6cm]{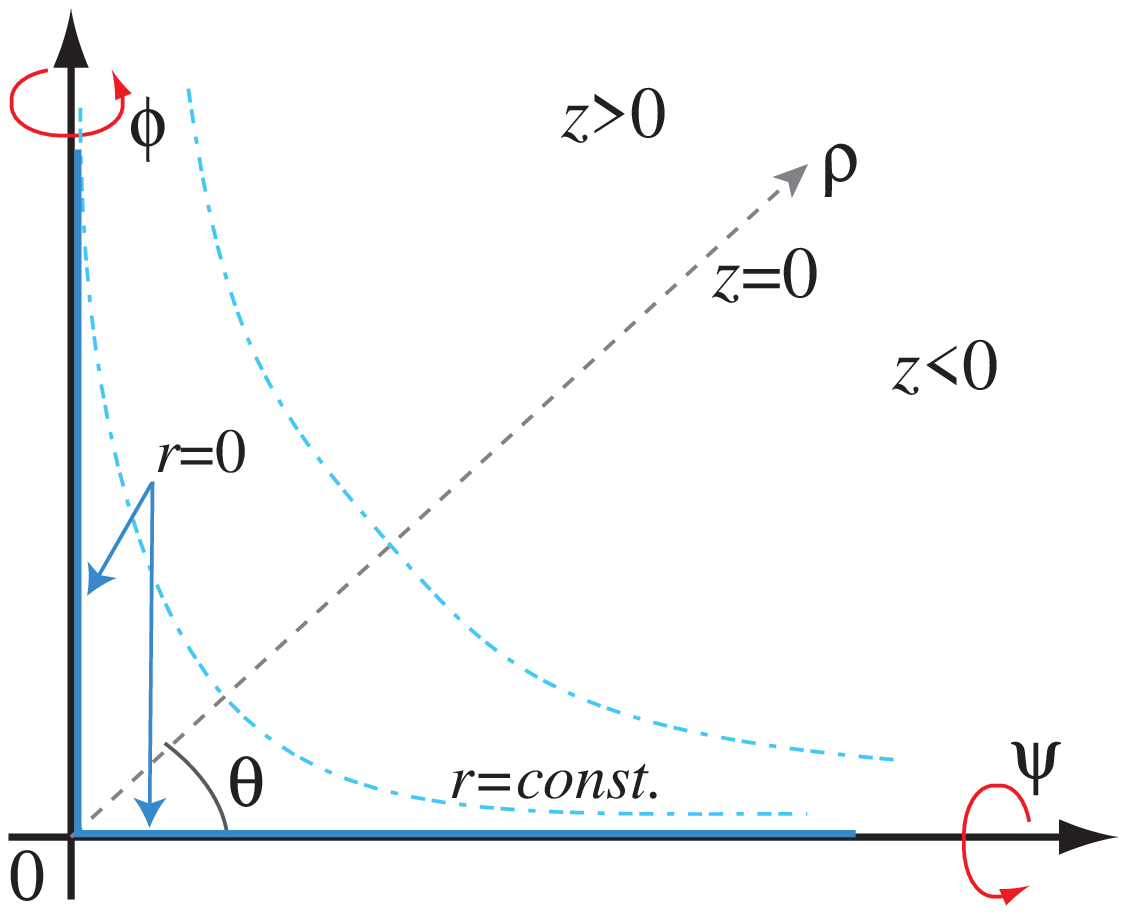}
\hspace{1cm}
\includegraphics[width=6.5cm]{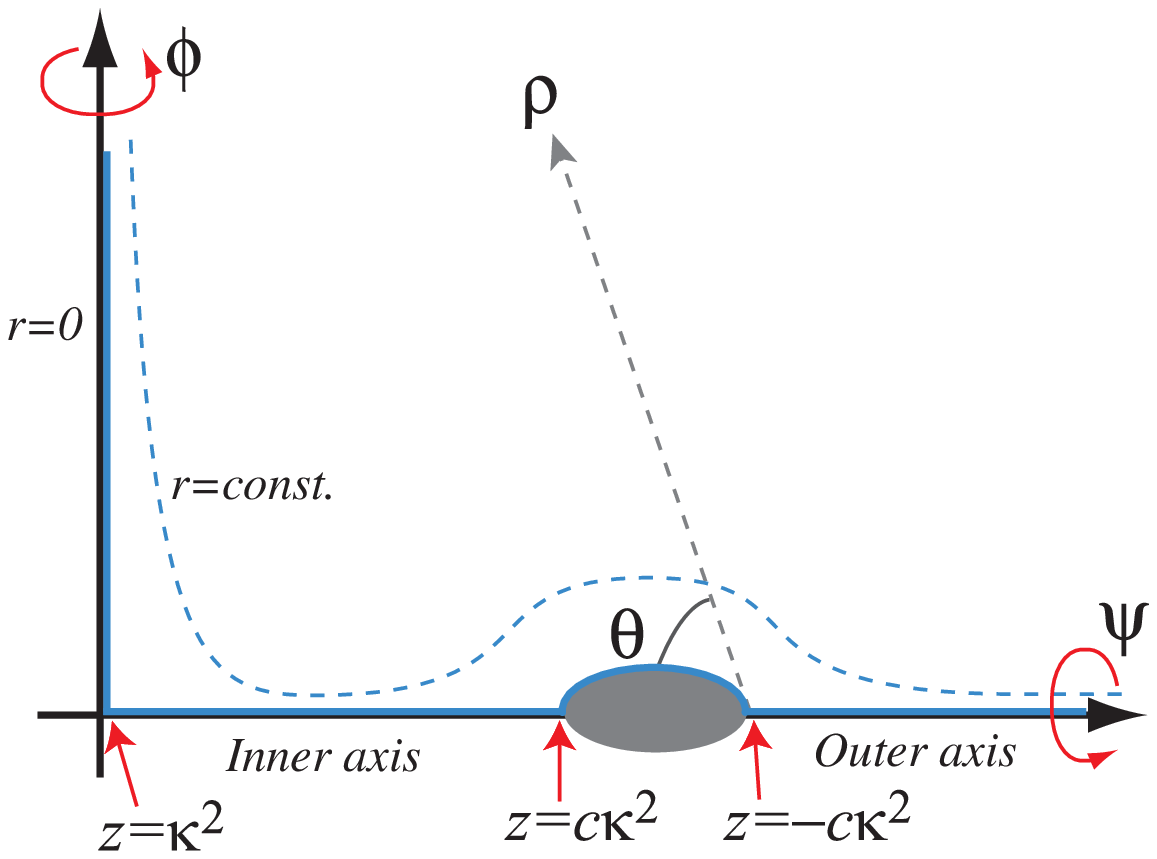}
\end{center}
\caption{
Canonical coordinates for Minkowski spacetimes (\ref{eq:5D Mink canonical}) and black rings (\ref{eq:Emparan-Reall sol}).
\label{fig:rz-coordinates}
}
\end{figure}

Putting $x^1=t$, $x^2=\phi$, $x^3=\psi$ where both $x^2$ and $x^3$ are periodic coordinates, the five-dimensional Minkowski spacetimes is 
\begin{eqnarray}
&&
ds^2  = 
  - dt^2 + (\sqrt{r^2+z^2} - z) d\phi^2
  + ( \sqrt{r^2+z^2} + z) d\psi^2
  + \frac{\epsilon^2}{2\sqrt{r^2+z^2}} (dr^2+dz^2).
\label{eq:5D Mink canonical}
\end{eqnarray}
The constant parameter $\epsilon$ is related to the periodicity of $\phi$ and $\psi$, and they are given by $\Delta \phi = \Delta \psi = 2\pi \epsilon$.
Appropriate rescalings of $r,z$ and the angular coordinates allows us to take $\epsilon=1$, and it is the standard form of the Minkowski spacetimes without conical singularity. However, the above form of Minkowski metric will be useful when we consider the spacetimes of black rings in which the periodicity is not necessary to be $\Delta \phi = \Delta \psi = 2\pi$.
The coordinate system is illustrated in Fig.~\ref{fig:rz-coordinates}.

In five-dimensions, the exact solution of rotating BRs can be expressed in the canonical form of metric. 
The spinning black ring is given by
\begin{eqnarray}
&&  g_{tt} = - \frac{W_-}{W_+},
\qquad
g_{\phi \phi}=   - \frac{r^2}{g_{tt}g_{\psi\psi}} + \frac{g_{t\phi}^2}{g_{tt}},
\qquad
  g_{\psi \psi}=  \frac{ (R_3 + z -\kappa^2 ) (R_2 -z + c \kappa^2)}
                     { R_1 - z -c \kappa^2 } ,
\nonumber
\\
&&
 g_{t\phi} 
= 
   - \frac{4c\kappa \sqrt{(1+c)}}{ \sqrt{1-c}  }
          \frac{ R_3 -R_1 + (1+c)\kappa^2  }{ W_+ } ,
\qquad
  e^{2\nu} =  W_+ 
     \frac{ (1-c) R_1 + (1+c) R_2 + 2c R_3 }{8 (1-c^4) R_1 R_2 R_3},
\label{eq:Emparan-Reall sol}
\end{eqnarray}
and $g_{\phi\psi}=g_{t\psi}=0$, where we have introduced 
\begin{eqnarray}
&& R_{1} \equiv \sqrt{r^2 + (z+c \kappa^2)^2}, 
\quad
 R_{2} \equiv \sqrt{r^2 + (z-c \kappa^2)^2}, 
\quad
 R_{3} \equiv \sqrt{r^2 + (z-  \kappa^2)^2}, 
\nonumber
\\
&&  W_\pm \equiv (1+c) R_1 + (1-c) R_2 - 2c R_3 \pm 4 c \kappa^2 , 
\nonumber
\end{eqnarray}
Here, $0<c<1$ and $\kappa>0$ are parameters.
Roughly speaking, the parameter $\kappa$ sets the scale for the solution so that it measures the radius of the ring. $c$ measures the ratio between the horizon radius of $S^2$ and the radius of the ring. So smaller values of $c$ corresponds to thinner ring.
In the limit of $c\to 0$ and $\kappa \to 0$, the Minkowski metric is recovered in the form (\ref{eq:5D Mink canonical}).
The BRs have no conical singularity if the periodicity of angles takes
\begin{eqnarray}
&&  \Delta \phi 
    = \Delta \psi 
    \equiv 2\pi \epsilon_{(bg)}
    = \frac{2\pi}{ \sqrt{1+c^2}}.
\end{eqnarray}

\subsection{Rod-structure}
\label{subsec:rod-structure}

The stationary axisymmetric solutions expressed in the canonical form of metric can be uniquely characterized by its boundary condition on the $z$-axis, known as rod-structure. 
The rod-structure can be rephrased as physical requirements/conditions of a black hole solution, so that it plays an important role to find a new solution.

Let us briefly discuss what the rod-structure is. 
The $z$-axis can be divided into the $n+1$ intervals, $[-\infty, z_1], [z_1, z_2], \cdots, [z_{n}, a_\infty]$. 
A necessary condition for a regular solution is that precisely one eigenvalue of $g_{ab}(0,z)$ becomes zero for a given $z$, except at the isolated points between the intervals.
The $n+1$ intervals are called rods of the solution $g_{ab}$, and the rod-structure for a solution is defined by
\begin{eqnarray}
  \lim_{r \to 0} g_{ab}(r,z_k) v^b_{(k)}=0 ,
\label{eq: def g*r=0}
\end{eqnarray}
with non-vanishing vectors $v_{(k)}$ $(k=1,\cdots,n)$, on the $k$-th interval (rod). 
Each vector associated to a rod has its own direction. 
If $\lim_{r \to 0} g_{ab}v^av^b /r^2$ is negative (positive), the direction of the rod is timelike (spacelike). 
A timelike rod corresponds to a horizon, while a spacelike rod corresponds to a compact direction. 
Moreover, a semi-infinite spacelike rod corresponds to an axis of rotation, and the associated coordinate is a rotation angle. 
These properties of rod and direction provide useful information of spacetimes.

Taking a closer look at the EOMs near $r=0$, we find that a solution $g_{ab}$ behaves as $g_{ab} = O(1) + O(r^2) + \cdots$, where the coefficient of each term will be a function of $z$. 
Except the isolated points between the rods, 
$g_{ab}$ stay finite at $r=0$ in general, but some metric components  vanishes according to the rod structure. 
Suppose $g_{11}$ behaves as $g_{11} \approx n(z) r^2 $ on some rod.
Then, from $R_{rz}=0$, we find 
\begin{eqnarray}
 e^{2\nu} = \gamma^2 n(z) + O(r^2)
\end{eqnarray}
to leading order for $r=0$, where $\gamma$ is a constant.
Then the relevant part of metric is 
\begin{eqnarray}
   ds^2 \sim  n(z) [ r^2 (dx^1)^2 + \gamma^2  dr^2 ]. 
\end{eqnarray}
If $x^1$ is a spacelike coordinate, a conical singularity appears on the axis, unless $x^1$ is a periodic coordinate with period 
\begin{eqnarray}
  \Delta x^1 = 2\pi\gamma.
\end{eqnarray} Therefore, for a rod in a spacelike direction, a specific periodicity is required to avoid a conical singularity. 
The same argument also holds for a timelike coordinate.  
If $x^1$ is a timelike coordinate, there is a horizon at $r=0$. By a Wick rotation, the Euclidean time must be periodic and it means that the temperature $T$ associated to the horizon is  
$ T = {1}/{2\pi \gamma}$. 
For stationary case, however, the direction $v^b$ will have multiple non-vanishing components and we have to take into account it. 
Furthermore, the BRs have no Euclidean counterpart that is real without conical singularity.

\section{Formulation of doubly spinning black rings}
\label{sec:Formulation of doubly spinning black rings}

The canonical coordinates have rectangular boundary, and they are suited for numerics.
Under the canonical form of metric, the field equations are non-linear coupled equations with elliptic second-order partial differential operator. 
Besides, the boundary conditions are well imposed by the rod-structure, once we fix the type of black holes. Thus we will be able to apply a numerical method for stationary cases. 
In this section, we formulate the black ring with two spins and discuss the necessary boundary conditions, including the basic strategy to solve the equations.

To deal with general black rings carrying two spins, we define the general metric in the following form; 
\begin{align}
     e^{2\nu} &= e^{2\nu^{(bg)} } M_{\nu}  , 
& \quad
 g_{\phi\phi} &= g_{\phi\phi}^{(bg)} M_{\phi\phi},
\quad 
&    g_{\psi\psi} &= g_{\psi\psi}^{(bg)} M_{\psi\psi},
\nonumber
\\
  g_{t\phi} &= g_{t\phi}^{(bg)} M_{t\phi}M_{\phi\phi} ,
\quad
&  g_{t\psi} &= g_{t\psi}^{(bg)} M_{t\psi}M_{\psi\psi} ,
\quad
&  g_{\phi\psi} &= M_{\phi\psi}.
\label{eq:general ansatz for 5D case}
\end{align}
Although we define $g_{tt}=g_{tt}^{(bg)}M_{tt}$, it is solved from  the algebraic relation (\ref{eq:r2=detg}). 
$g_{t\psi}^{(bg)}$ is defined by
\begin{eqnarray}
    g_{t\psi}^{(bg)} = - \left(\frac{2c \kappa^2}{R_1+R_2}\right)^2 g_{\psi\psi}^{(bg)}. 
\end{eqnarray}
Here, $M_{ab}(r,z)$ and $M_\nu(r,z)$ are unknown functions which we solve. 
We adopt the Emperan-Reall's solution (\ref{eq:Emparan-Reall sol}) as the ``background" metric $g_{ab}^{(bg)}$, so that the Emparan-Reall's solution is recovered in the limit of
\begin{eqnarray}
&& M_{t\phi}, M_{\phi\phi}, M_{\psi\psi}, M_{\nu} \to 1,
\nonumber
\\
&& M_{t\psi}, M_{\phi\psi}  \to 0.
\label{eq:limint Mab to ER}
\end{eqnarray}
Our metric ansatz, in particular $g_{t\phi}$ and $g_{t\psi}$, are chosen in order to yield simple expression for physical quantities, such as the angular velocity of horizon. 
Besides, as we will see below, an advantage of adopting the above ansatz is that the full metric will automatically satisfy the desired rod-structured as long as $M_{ab}$ are smooth functions.

From Eqs. (\ref{eq:R_ab}) and  (\ref{eq:R_zz}), the coupled non-linear PDEs for the functions $M_{ab}$ are symbolically written as 
\begin{eqnarray}
   \nabla^2 M = S(M), 
  \label{eq:symbolic eq DDM=S}
\end{eqnarray}
with the non-linear source term $S$, where  $M = M_{ab}$ or $M_{\nu}$. 
To solve this system as a boundary value problem, we specify the boundary conditions by explicitly analyzing the rod structure and the asymptotic behavior for the assumed general metric (\ref{eq:general ansatz for 5D case}).

\subsection{Asymptotic analysis}
\label{subsec:Asymptotic analysis}

Let us first specify the asymptotic behavior of the solution. 
In the following new coordinates,
\begin{eqnarray}
  r =   {\rho}^{2} \sin \theta ,
\quad
  z + c \kappa^2 = {\rho}^{2} \cos \theta,
\label{eq:rz-spheroidal def2}
\end{eqnarray}
the metric asymptotes to the five-dimensional Minkowski space (\ref{eq:5D Mink canonical}) in the limit of $\rho \to \infty$. 
Assuming the general periodicity of $\Delta \phi = \Delta \psi = 2\pi \varepsilon$, the subleasing order terms are found to be  
\begin{align}
\delta g_{tt}   & \simeq  \frac{4M}{3\pi \epsilon^2 \rho^{2}}, 
\quad
& \delta g_{\phi\phi} & \simeq  \frac{4(M+\eta) \sin^2 (\theta/2)}{3\pi\epsilon^2},
\quad
& \delta g_{\psi\psi} & \simeq  \frac{4(M-\eta) \cos^2(\theta/2)}{3\pi \epsilon^2}, 
\nonumber
\\
  \delta g_{t\phi} & \simeq  -\frac{2 J_\phi \sin^2 (\theta/2)}{\pi \epsilon^3  \rho^{2}}, 
\quad
& \delta g_{t\psi} & \simeq  -\frac{2 J_\psi \cos^2 (\theta/2)}{\pi \epsilon^3  \rho^{2 }},
\quad
& \delta g_{\phi\psi} & \simeq  - \frac{ \zeta \sin^2 \theta }{\epsilon^4  \rho^{2 }}. 
\label{eq:asympt5d}
\end{align}
Here, $M$ and $J$ are mass and angular momentum, respectively, and $\zeta$ and $\eta$ are constants. 
The periodicity $\epsilon$ is related to the leading order of $e^{2\nu}$, which is given by
\begin{eqnarray}
&& e^{2\nu}
   \simeq  \frac{\epsilon^2}{2 \rho^{2}} + O(1),
\label{eq:asym e2nu}
\end{eqnarray}

From these behaviors and those of background metric (\ref{eq:Emparan-Reall sol}), the metric variables $M_{ab}$ must obey the following asymptotic decays, 
\begin{eqnarray}
&&
   M_{ab} \sim  1 + \frac{\alpha_{ab}}{ \rho^{2}} ,
  \qquad 
  (M_{tt},M_{\phi\phi},M_{\psi\psi})
\nonumber
\\
&& 
  M_{ab} \sim \alpha_{ab} + O(\rho^{-2}) ,
  \qquad  
  (M_{t\phi}, M_{t\psi}, M_{\nu})
\qquad\qquad\quad
\cr
&&  
 M_{\phi\psi} \sim \alpha_{\phi\psi} \frac{\sin^2 \theta}{\rho^{2} } ,
\label{eq:define asymptotic behaviour M}
\end{eqnarray}
were $\alpha_{ab}$ are constant coefficients.
These asymptotic behaviors are necessary to be imposed as the Neumann boundary conditions at the outer boundary. 
From each coefficient, the physical quantities can be read off as 
\begin{eqnarray}
&& M = \frac{3\pi \epsilon^2}{4} \alpha_{tt},
\qquad
 M  = \frac{3\pi \epsilon^2}{2}\left[ 
                 \frac{ 4c \kappa^2}{(1-c)} 
            + \alpha_{\phi\phi} + \alpha_{\psi\psi}
                  \right],
\nonumber
\\
&& 
J_1 = \alpha_{t\phi}  \frac{2\pi \epsilon^3 \kappa^3  c (1+c)^{3/2}}{(1-c)^{3/2}},
\qquad
J_2 = \pi \epsilon^3 c^2 \kappa^4 \alpha_{t\psi}, 
\qquad
\epsilon^2 = \epsilon^2_{(bg)} \alpha_{\nu}
\end{eqnarray}
where $\alpha_{tt}$ yields  $\alpha_{tt} \to 4c \kappa^2/(1-c)$ in the limit of ${J_2 \to 0}$.

\subsection{Boundary conditions and Rod structure}
\label{subsec:Rod structure}

We analyze the boundary conditions of the general BR metric and specify the rod-structure. 
Since the geometry of the general BRs with two spins is the same as the black ring with single spin, the basic rod-structure should be also the same. 
The rod-structure is classified into four parts: the semi-infinite spacelike rod $[-\infty, -c \kappa^2]$, finite timelike rod $[-c \kappa^2, c \kappa^2]$, finite spacelike rod $[+c \kappa^2, \kappa^2]$, and the semi-infinite spacelike rod $[\kappa^2, \infty]$. 
Near the $r=0$, the metric components behave as $g_{ab} = O(1) + O(r^2)$, so that the basic boundary conditions are the Neumann boundary condition:
\begin{eqnarray}
    \partial_r  g_{ab} =0.  \quad (r=0)
\end{eqnarray}
In addition to this condition, some types of Dirichlet boundary conditions are necessary. 
An important point is that the boundary conditions, which can be translated into the rod-structure, is determined by requiring physical (or regularity) conditions on the $z$-axes.

\begin{itemize}

  \item  
The semi-infinite spacelike rod $[\kappa^2, \infty]$ corresponds to the $\phi$-axis. 
This means that 
\begin{eqnarray}
 g_{\phi\phi}=g_{t\phi}=g_{\phi\psi}=0 
\label{eq:phi-axis g=0}
\end{eqnarray}
should be satisfied there.
This physical requirement is equivalent to having a rod in the direction $v=(0,1,0)$, i.e., the $\partial/\partial x^2$ direction.  
From the property of background metric, $g_{\phi\phi}$ and $g_{t\phi}$ always satisfy the above condition.
The non-trivial condition is 
\begin{eqnarray}
    M_{\phi\psi}=0.
\end{eqnarray}

As discussed in Sec.~\ref{subsec:rod-structure}, the periodicity of angular coordinate is given by
$
\Delta \phi = 2\pi \lim_{r\to 0} \sqrt{ {r^2 e^{2\nu}}/{g_{\phi\phi}}}, 
$
 which yields
\begin{eqnarray}
 \Delta \phi = \frac{2\pi}{\sqrt{1+c^2}} \sqrt{ \frac{M_{\nu}}{M_{\phi\phi}}}. 
    \quad (\text{$\phi$-axis})
\label{eq:Dphi on phi-axis}
\end{eqnarray}
Since the periodicity does not change along the axis, the above result means 
$M_{\nu}/M_{\phi\phi} = \mathrm{const.} $ on the axis. 
In the asymptotic limit ($\rho\to 0$), we obtain $M_{\phi\phi} \to 1$ and  $M_\nu \to \text{const.}$, and so the asymptotic value of $M_{\nu}$ determines the periodicity (see Eq.~(\ref{eq:asym e2nu})).
The condition (\ref{eq:Dphi on phi-axis}) can be used as a Dirichlet condition for $M_\nu$, once its asymptotic value is fixed.

  \item  
Similarly, the semi-infinite spacelike rod $[c\kappa^2, \kappa^2]$ corresponds to the axis of $\psi$ which locates inside the ring. 
We call this portion of axis the inner axis. 
In this case, the physical requirement (and the rod structure) gives  
\begin{eqnarray}
   g_{\psi\psi} = g_{t\psi} = g_{\phi\psi} =0 , 
\label{eq:psi-axis g=0 1}
\end{eqnarray}
and we find $ M_{\phi\psi} =0$. 
The direction of the rod is $v=(0,0,1)$.
Note that the background metric $g_{t\psi}^{(bg)}$ is chosen so as to satisfy this condition automatically. 
With the same argument above, the periodicity of this axis is found to be
\begin{eqnarray}
 \Delta \psi  = \frac{2\pi}{\sqrt{1+c^2}} \sqrt{ \frac{M_{\nu}}{M_{\psi\psi}}},
    \quad (\text{$\psi$-axis})
\label{eq:Dpsi on psi-axis innter}
\end{eqnarray}
to avoid a conical singularity. 
This condition can be also used as a Dirichlet boundary condition for $M_\nu$.

\item 

The finite timelike rod can be put at $[-c\kappa^2, c\kappa^2]$, and it is the horizon. 
From (\ref{eq: def g*r=0}) with $v$ being the vector 
$v = (1,\omega_\phi, \omega_\psi)$, we find that the angular velocities on the horizon are 
\begin{eqnarray}
 \omega_{\phi} =  \frac{\sqrt{1-c}}{2\kappa \sqrt{1+c}} M_{t\phi},  
\quad
 \omega_{\psi} = M_{t\psi}. 
\label{eq:def omega2,3}
\end{eqnarray}
Here we have required 
\begin{eqnarray}
  M_{\phi\psi}=0. 
\label{eq:m23=0 at horizon}
\end{eqnarray}
The physical angular velocities are given by 
\begin{eqnarray}
  \Omega_{\phi} =  
   \frac{2\pi}{\Delta \phi} \omega_\phi,
\quad
  \Omega_{\psi} =  
   \frac{2\pi}{\Delta \psi} \omega_\psi.
\end{eqnarray}
As we notice, $\Omega_\phi$ and $\Omega_\psi$ are constant over the horizon as long as $M_{t\phi}$ and $M_{t\psi}$ are constant. 
We will be able to use these constants as parameters of specifying a solution. This is the reason why we have adopted the metric ansatz in the form of (\ref{eq:general ansatz for 5D case}).
Note that physically the rod-structure on the horizon is equivalent to finding a null Killing vector 
$\xi^a = (\partial_t)^a + \omega_\phi (\partial_\phi)^a + \omega_\psi (\partial_\psi)^a$, where $\omega_\phi$ and $\omega_\psi$ are defined by $\omega_b = - g_{tb}/g_{bb}$ with $g_{\phi\psi}=0$.

From the generator of the Killing horizon, the temperature is 
\begin{eqnarray}
&& T = \frac{1}{2\pi \gamma}, 
\label{eq:T and S}
\\
&& \gamma 
\equiv
  \frac{ 4c \kappa 
    \sqrt{M_{\phi\phi} M_{\psi\psi} M_{\nu} }  }{(1-c) \sqrt{(1+c^2)}}
    \biggr|_{r=0}.
\nonumber
\end{eqnarray}
Here, $1/\gamma$ is the surface gravity. 
From the constancy of surface gravity on the horizon, 
$M_{\phi\phi} M_{\psi\psi} M_{\nu}|_{r=0}$ is constant over the horizon. 
Then the entropy is given by 
\begin{eqnarray}
&& S = 
   \frac{ 2c^2 \kappa^3 \Delta \phi \Delta\psi }{(1-c) \sqrt{1+c^2}} 
   \sqrt{ M_{\phi\phi} M_{\psi\psi} M_{\nu}    }.
\end{eqnarray}
Note that the entropy and temperature follow the Smarr relation and the first law with other physical quantities:
\begin{eqnarray}
&&  \frac{2}{3} M =  T S + \sum_i \Omega_i J_i , 
\qquad
    dM = TdS + \sum_i  \Omega_i dJ_i .
\label{eq:Smarr}
\end{eqnarray}

\item

The infinite spacelike rod $[-\infty, -c \kappa^2]$ becomes the axis of $\psi$ which locates outside the ring. 
The direction of the rod is $v=(0,0,1)$, which means
\begin{eqnarray}
   g_{\psi\psi}=g_{t\psi} = g_{\phi\psi} =0.
\label{eq:psi-axis g=0 2}
\end{eqnarray}
Again, we obtain a non-trivial condition $M_{\phi\psi}=0$. 
To avoid a conical singularity on this axis, $\psi$ needs to have the period 
\begin{eqnarray}
\Delta \psi 
=
  \frac{2\pi}{\sqrt{1+c^2}} \sqrt{ \frac{M_{\nu}}{M_{\psi\psi}}}.
\label{eq:delta psi at outer axis}
\end{eqnarray}

\end{itemize}

The above analysis of rod-structure gives us additional information on the boundary conditions, and all of these conditions are sufficient to solve the Einstein equation as a boundary value problem. 
Our choice of the background metric $g^{(bg)}_{ij}$ is chose so that the full metric automatically satisfy the Dirichlet boundary conditions (\ref{eq:phi-axis g=0}), (\ref{eq:psi-axis g=0 1}), (\ref{eq:psi-axis g=0 2}) on the axis except for $g_{\phi\psi}$, in addition to providing simple expression for the rotational velocity on the horizon. 
The difference in the rod-structure between the BR with single spin and the doubly spinning BR is the direction of the vector $v$ on the horizon.

\section{Perturbative approach}
\label{sec:Perturbative scheme}

\subsection{First order perturbation}

We are ready to solve the field equations as a well-defined problem.
Although it is very challenging to develop an elaborate numerical code that solves the nonlinear equation (\ref{eq:symbolic eq DDM=S}) directly, we discuss a method to solve the system perturbatively. 
Passing this approach, we can confirm the validity of the formulation and the specification of the rod-structure discussed so far.

Based on the behavior (\ref{eq:limint Mab to ER}) near the background metric, the metric functions are expanded as 
\begin{eqnarray}
&& M_{ab} = m_{ab}^{(bg)} 
            + \sum_{n=1}^{\infty} \tilde{\omega}^n_\psi m_{ab}^{[n]}, 
\label{eq:def pert m_ab}         
\end{eqnarray}
where $\tilde{\omega}_\psi$ is an expansion parameter and the first term is $m_{ab}^{(bg)}= 0$ or $1$.
For instance, 
\begin{eqnarray}
M_{\phi\phi} &=& 1+ \tilde{\omega}_\psi m_{\phi\phi}^{[1]} + \cdots,
\\
M_{\phi\psi} &=&  \tilde{\omega}_\psi m_{\phi\psi}^{[1]} + \cdots.
\end{eqnarray}
For brevity we will omit hereafter the superscript $[n]$ denoting the order of perturbations whenever it is obvious. 
From (\ref{eq:def omega2,3}) (on the horizon), the expansion parameter $ \tilde{\omega}_\psi$ is just the second angular velocity on the horizon at the leading order, 
\begin{eqnarray}
\omega_\psi \approx \tilde{\omega}_\psi, 
\end{eqnarray}
and we do not need to distinguish them hereafter. 
This is the most natural expansion parameter in the present issue because the angular velocity vanishes in the limit of black rings with single spin and will remain small finite value even for maximum rotation.

At the first order of $\omega_\psi$, the coupled equations are symbolically written as
\begin{eqnarray}
&&  \mathcal{P} \left( 
  \begin{array}{c}
    m_{t\psi}   \\
    m_{\phi\psi}   \\
  \end{array}
\right) = 0, 
\label{eq:eqs for m13m23}
\\
&&  \nabla^2 m_{\psi\psi} = 0,
\label{eq:eqs for m33 1st}
\\
&&
  \mathcal{Q} \left( 
  \begin{array}{c}
    m_{\phi\phi}   \\
    m_{t\phi}    \\
  \end{array}
\right) = S(m_{\psi\psi}),
\label{eq:eqs for m11 m12 1st}
\\ 
&&  \nabla^2 m_{\nu}  = S(m_{\phi\phi}, m_{t\phi}, m_{\psi\psi}).
\end{eqnarray}
Here $S$ is a source term depending on variables $m_{ab}$. 
$\mathcal{P}$ and $\mathcal{Q}$ represent a general second-order elliptic operator, e.g.,
\begin{eqnarray}
 \mathcal{P} \equiv  A \partial_r^2 + B \partial_r + C \partial_z^2 
     + D \partial_z + E, 
\end{eqnarray}
where the coefficients $A, B, \cdots$ are $2\times 2$ matrices of functions of $\{r,z\}$, which contain background metric. 
It is remarkable that the equations for $\{ m_{t\psi}, m_{\phi\psi} \}$ decuples from the others. 
Thus by solving (\ref{eq:eqs for m13m23}), we can introduce the second angular momentum into the BRs with single spin. 
Eq. (\ref{eq:eqs for m33 1st}) is solved analytically, and one finds that all non-trivial solutions do not satisfy the asymptotic boundary condition. Hence, we have $m_{\psi\psi}=0$, and Eq. (\ref{eq:eqs for m11 m12 1st}) is also decoupled from the others.

The equation (\ref{eq:eqs for m11 m12 1st}) with $m_{\psi\psi}=0$ has a regular solution. 
The angular velocity $\omega_\phi \propto m_{t\phi}$ on the horizon is a constant parameter, and once it is provided, the solution is uniquely determined.
In fact, if we set $m_{t\phi} = 0$ on the horizon, only the trivial solution is found numerically. 
Note that the operator $\mathcal{Q}$ are invariant under a constant shift of $m_{\phi\phi} \to m_{\phi\phi} + \text{const.}$,  $m_{t\phi} \to m_{t\phi} - \text{const}$. This constant mode will be fixed by the asymptotic condition $m_{\phi\phi} \to 0$.

The meaning of the independent perturbations of (\ref{eq:eqs for m11 m12 1st}) is clearly that the mass and angular momentum of the background metric can be changed infinitesimally irrespective of the second spin. This is consistent with the continuous sequence of BRs with single spin. 
We are not interested in these modes so that we turn off these perturbations.  
In this case, $m_{\nu}$ has no regular solution except a constant mode, and so we also kill this perturbation. 
After all, we take 
\begin{eqnarray}
    m_{\psi\psi}^{[1]} = m_{\phi\phi}^{[1]} = m_{t\phi}^{[1]} = m_{\nu}^{[1]} = 0,
\end{eqnarray}
at the linear order.

\begin{figure}[t]
\begin{center}
\includegraphics[width=7cm]{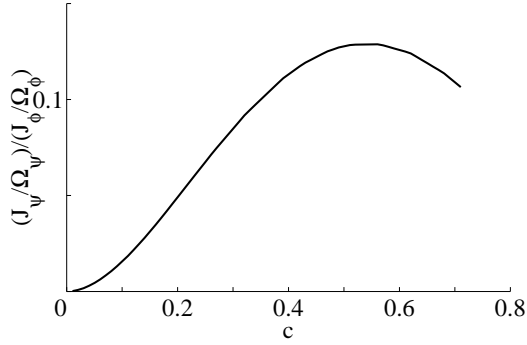}
\end{center}
\caption{
\label{fig:j2overj1}
Plot of  $({J_\psi}/{\Omega_\psi})({J_\phi}/{\Omega_\phi})^{-1}$. 
}
\end{figure}

Even with these simplifications, the above PDEs for the perturbations do have complicated coefficients which are functions of $\{r, z\}$, and it is impossible to carry out separation of variables, except for $m_{\psi\psi}$ and $m_\nu$. 
We solve the PDEs by employing a numerical method.
An example of numerical solutions is shown in Fig.~\ref{fig:m13,m23} for a particular set of $c$ and $\kappa$.
From a sequence of numerical solutions, we read off the angular momentum $J_\psi$. 
In Fig.~\ref{fig:j2overj1}, we show how the second perturbative angular momentum changes with respect to the background parameter $c$.
To eliminate the expansion parameter $\omega_\psi$ and $\kappa$, we combine the following two quantities at the leading order,
\begin{eqnarray}
\frac{J_\phi}{\Omega_\phi} &=& 
        \frac{4c(1+c)^2 \pi \kappa^4}{(c^3-c^2+c-1)^2},
\cr
\frac{J_\psi}{\Omega_\psi} &=& \frac{\pi c^2\kappa^4}{(1+c^2)^2} \alpha_{t\psi}.
\end{eqnarray}
Figure~\ref{fig:j2overj1} shows that the second angular momentum $J_\psi$ increases with respect the background angular momentum $J_\phi$ as $c$ increased until $c\lesssim 1/2$ (thin black rings).
Notice that the background angular momentum $J_\phi$ (and $J_\phi/\Omega_\phi$) increases monotonically as $c\to 1$ (limit of thick black rings).
Then the above result means that for a fixed angular velocity $\Omega_\psi$ the perturbative contribution of $J_\psi$ becomes relatively larger as $c$ increased. 
For the branch of fat BRs ($c \gtrsim 1/2$), $J_\phi/\Omega_\phi$ diverges in the limit of $c\to 1$, and so the relative contribution of $J_\psi/\Omega_\psi$ decreases as long as $J_\psi/\Omega_\psi$ is finite.

\begin{figure}[t]
\begin{center}
\includegraphics[width=7cm]{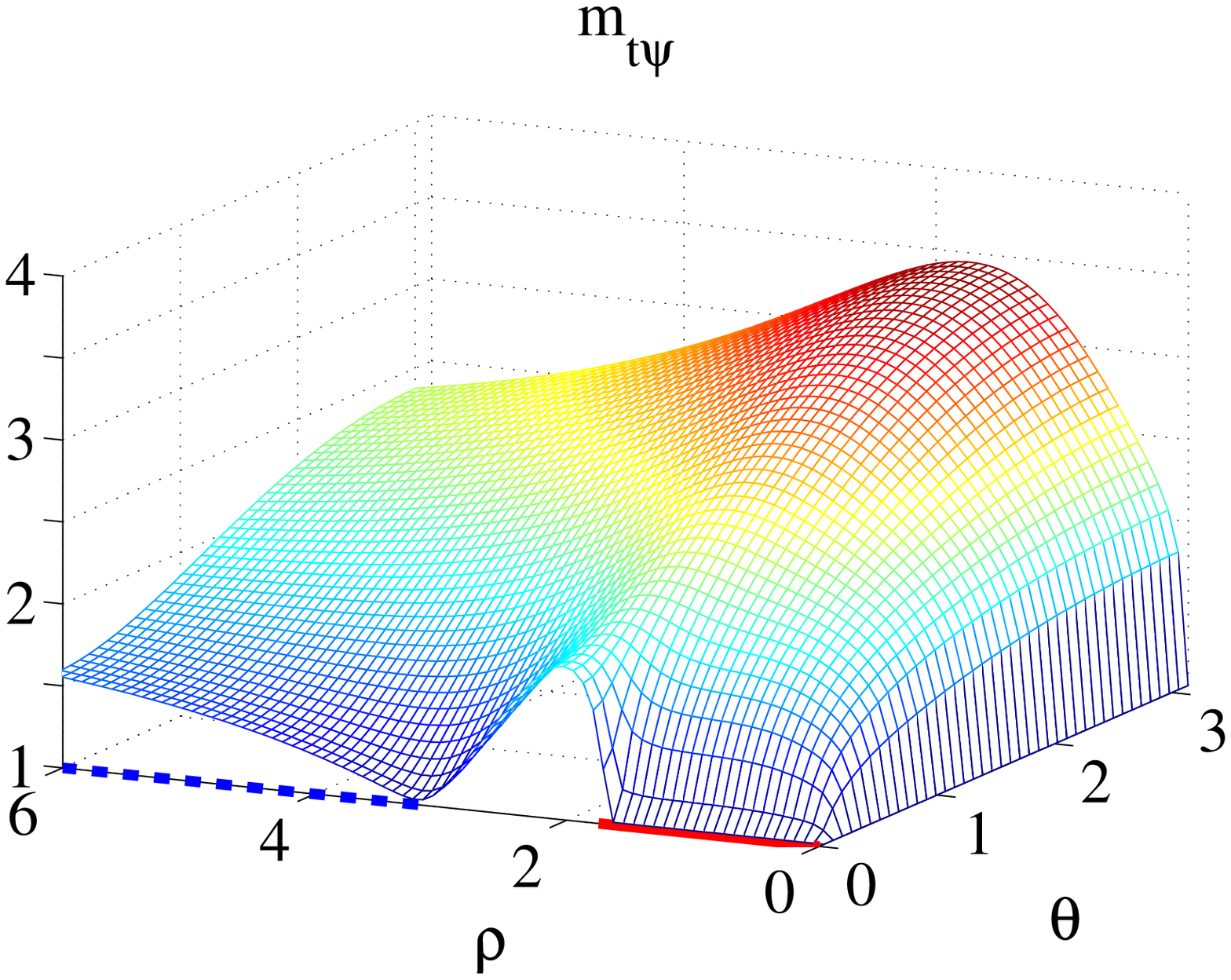}
\hspace{1cm}
\includegraphics[width=7cm]{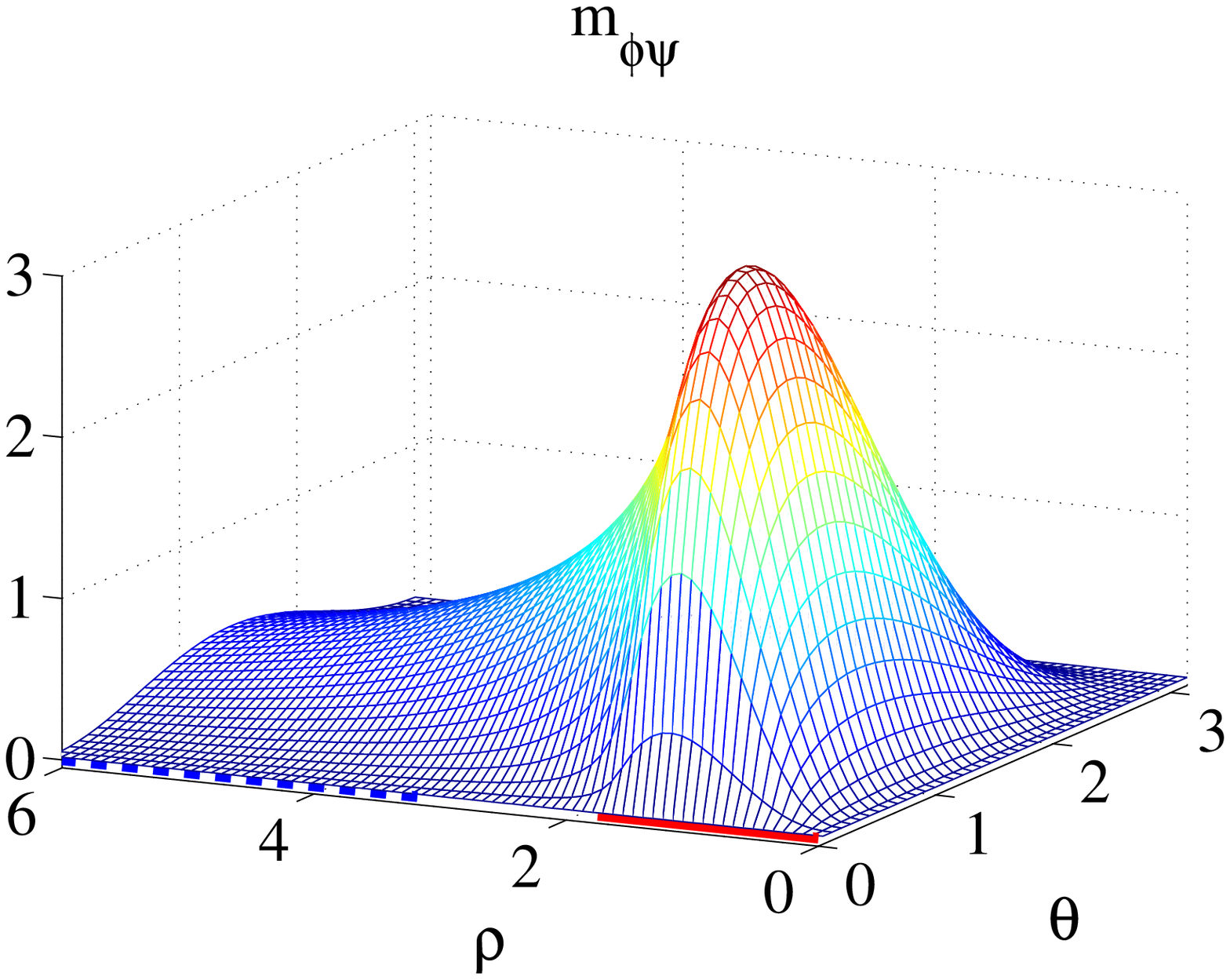}
\end{center}
\caption{
\label{fig:m13,m23}
Numerical solution of $m_{t\psi}$ and $m_{\phi\psi}$ in the coordinates (\ref{eq:rz-spheroidal def2}) for $c=0.18$ and $\kappa=2.9$. 
The thick and dashed lines on the $\rho$-axis ($\theta=0$) represent the horizon and $\phi$-axis in the $\{r,z\}$ coordinates, and the portion between them corresponds to the inner axis. 
The outer axis is at $\theta=\pi$. 
The asymptotic boundary is at $\rho\approx 12$, which is not covered in the above figures. 
}
\end{figure}

\subsection{Second order perturbation}

It is interesting to consider the backreaction from the second angular momentum on the background BRs.
It will allow us to check the consistency of perturbations and will give useful information about the doubly spinning BRs. 
The relevant Einstein equations at the second order are 
\begin{eqnarray}
 &&  \nabla^2  m_{\psi\psi}^{[2]} = S(m_{t\psi}, m_{\phi\psi}),
\label{eq:m33 -2nd order}
\\
&&
   {\mathcal{Q}} \left( 
  \begin{array}{c}
    m_{\phi\phi}^{[2]}    \\
    m_{t\phi}^{[2]}    \\
  \end{array}
\right) =  S(m_{\psi\psi}, m_{t\psi}, m_{\phi\psi}  ), 
\label{eq:m22 -2nd order}
\\
&&  \nabla^2  m_{\nu}^{[2]}  = S(  m_{\phi\phi}, m_{t\phi}, 
                        m_{\psi\psi}, m_{t\psi}, m_{\phi\psi}),
\label{eq:mrr-2nd order}
\end{eqnarray}
where $S$ represents source terms which contain quadratic of linear perturbations and/or second order variables. 
The inhomogeneous solutions of these equations carry the backreaction of second angular momentum. 
These equations will be solved numerically in descending order.

There are subtle issues in (numerically) solving these equations. 
Let us first discuss the equation (\ref{eq:m22 -2nd order}). 
The Neumann condition $\partial_r m_{ab} = 0$ are the basic boundary conditions for $m_{\phi\phi}$ and $m_{t\phi}$ at $r=0$. 
The exception is $m_{t\phi}$ on the horizon, which is constant over the horizon.
Since $m_{\phi\phi}$ and $m_{t\phi}$ have the homogeneous solutions, the constant value $m_{t\phi}$ on the horizon can be set arbitrary by adding the homogeneous solution.
However, this ambiguity is removed by noticing that the periodicity on the $\phi$-axis and that on the $\psi$-axis should be the same, to achieve the balance between forces in the ring. 
Expanding Eqs. (\ref{eq:Dphi on phi-axis}) and (\ref{eq:Dpsi on psi-axis innter}), 
\begin{eqnarray}
&& \Delta \phi = \frac{2\pi}{\sqrt{1+c^2}} 
                \left[ 1 + \frac{\omega_\psi^2}{2} (m_\nu - m_{\phi\phi}) \right],
\cr
&& \Delta \psi = \frac{2\pi}{\sqrt{1+c^2}} 
                \left[ 1 + \frac{\omega_\psi^2}{2} (m_\nu - m_{\psi\psi}) \right].
\end{eqnarray}
Assuming the continuity of $m_\nu$ at the boundary between $\phi$- and $\psi$-axes, we obtain
\begin{eqnarray}
 m_{\phi\phi} = m_{\psi\psi} .  \quad (r=0,z= \kappa^2)
 \label{eq: m33=m22 condition}
\end{eqnarray}
This condition fix the overall amplitude of $m_{\phi\phi}$; the homogeneous solution for (\ref{eq:m22 -2nd order}) is superposed on a specific solution of (\ref{eq:m22 -2nd order}) so as to satisfy the above condition. 
With this condition for $m_{\phi\phi}$, the angular velocity $\omega_\phi \propto m_{t\phi}$ is determined automatically.

Next, we discuss about $m_{\nu}$. 
Let us recall that this variable is defined by
\begin{eqnarray}
&& e^{2\nu} = e^{2\nu_{(bg)}} M_\nu
= e^{2\nu_{(bg)}} ( 1 + \omega_\psi^2 m_\nu + \cdots),
\end{eqnarray}
and its asymptotic value is related to the periodicity parameter $\epsilon$, 
\begin{eqnarray}
&& m_\nu =  \tilde{\alpha}_\nu + O(\rho^{-2}),
\cr
&& \epsilon^2 = \epsilon^2_{(bg)} (1+  \omega_\psi^2  \tilde{\alpha}_\nu), 
\end{eqnarray}
where $\tilde{\alpha}_\nu$ is defined by $ \alpha_\nu = 1 + \omega_\psi^2 \tilde{\alpha}_\nu$. 
Although the solution for $m_\nu$ is easily found by inverting the operator $\vec{\nabla}^2$,  we encounter a technical deficit in the present approach whenever we adopt numerical solutions.  
The issue is how we fix a constant mode involved in $m_{\nu}$, which directly affects $\tilde{\alpha}_\nu$. 
We notice that it is free to add a constant to a particular solution of $m_{\nu}$, which corresponds to changing the amplitude of $M_{\nu}$, or adding a constant in $\nu$, 
\begin{eqnarray}
 \nu \to \nu + \mathrm{const.}
\end{eqnarray}
Adding a constant mode is completely consistent with all the boundary conditions for $e^{2\nu}$, and no boundary conditions (and the PDE) can fix it. 
The value of $m_\nu$ (or $\nu$) is related to the periodicity of axes, which enters into the physical quantities, and so it has a physical meaning.
However, as long as we solve the PDE numerically, we can add an arbitrary constant to each obtained solution. 
One might think that we are able to fix the ambiguity through the Smarr relation between physical quantities. 
But this does not work because the periodicity $\epsilon$ is factored out as an overall factor, and the Smarr relation does not care about the periodicity \cite{Emparan:2004wy}.

What is the origin of this problem? 
Suppose that we find a full analytic solution for the PDEs, and consider how we fix the amplitude of $e^{2\nu}$. 
The amplitude will be fixed by requiring that the periodicity of angles becomes, e.g., $2\pi$ in the (continuous) limit of Minkowski spacetimes. 
This is the only way we can actually do for the black rings with single spin. 
On the other hand, however, if we find solutions numerically, they constitute a discrete sequence of solutions and each numerical solution can carry its own (arbitrary) periodicity of $\Delta \phi = \Delta \psi$. 
Since the periodicity (or arbitrary added constants) has no relation with each other, we have no principle to fix it a priori.

Except for this problem, there is no other ambiguity in solving $m_{\nu}$. 
To fix the constant mode, we will set
$ \alpha_\nu=0$ when we solve $m_\nu$.
Once we fix the constant mode, the boundary condition on $r=0$ are just Dirichlet conditions derived from (\ref{eq:Dphi on phi-axis}), (\ref{eq:Dpsi on psi-axis innter}), (\ref{eq:T and S}) and (\ref{eq:delta psi at outer axis}), where we use the fact that the temperature and periodicity are constant.
For instance, 
Eqs. (\ref{eq:Dphi on phi-axis}) and (\ref{eq:Dpsi on psi-axis innter}) gives 
\begin{eqnarray}
&& m_{\nu}^{\alpha=0} = m_{\phi\phi}, \quad (\phi\mathrm{-axis})
\cr
&& m_{\nu}^{\alpha=0} = m_{\psi\psi},  \quad (\psi\mathrm{-axis})
\end{eqnarray}
where $m_{\nu}^{\alpha=0}$ is a solution with $\alpha_{\nu}=0$.

\begin{figure}[t]
\begin{center}
\includegraphics[width=7cm]{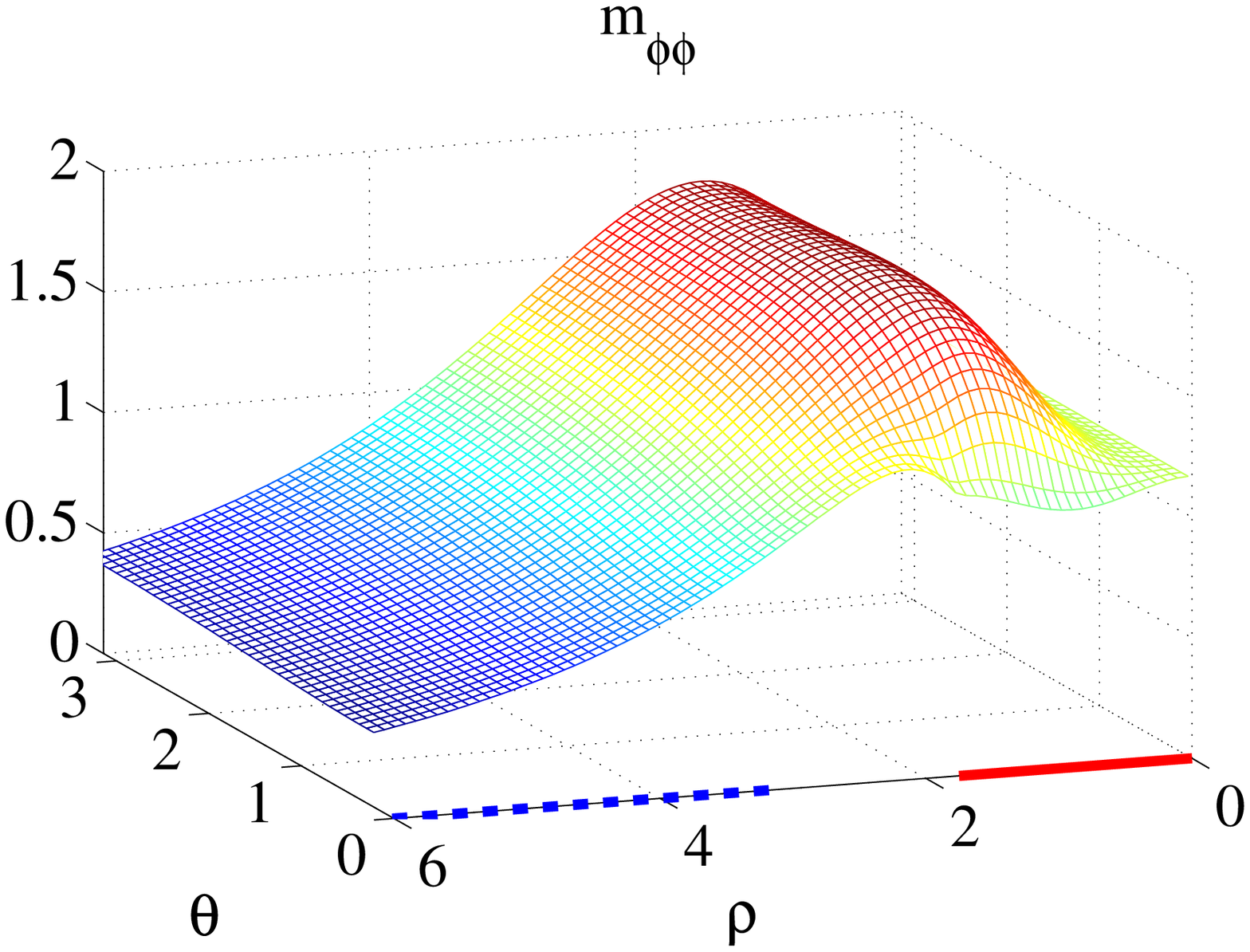}
\hspace{1cm}
\includegraphics[width=7cm]{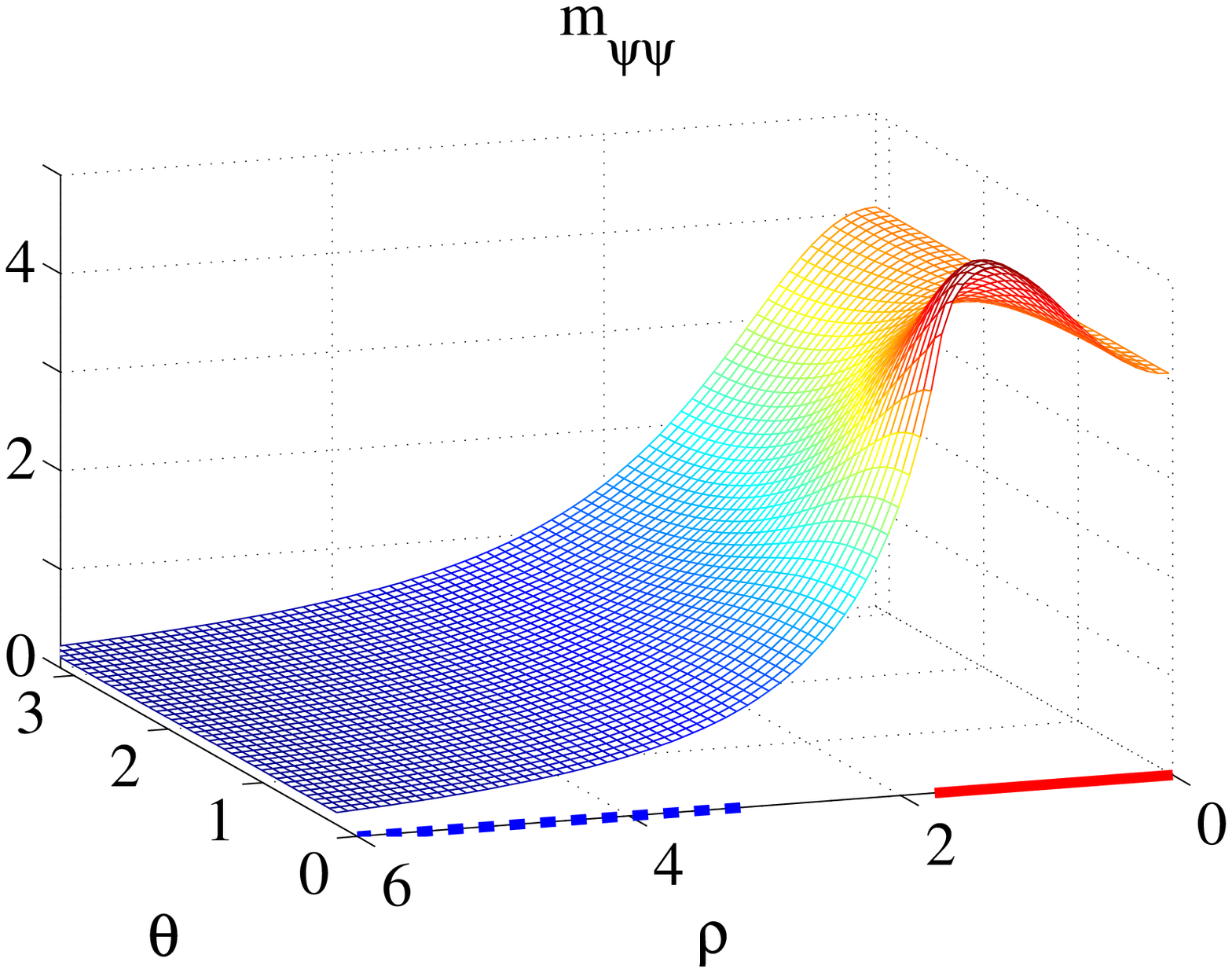}
\includegraphics[width=7cm]{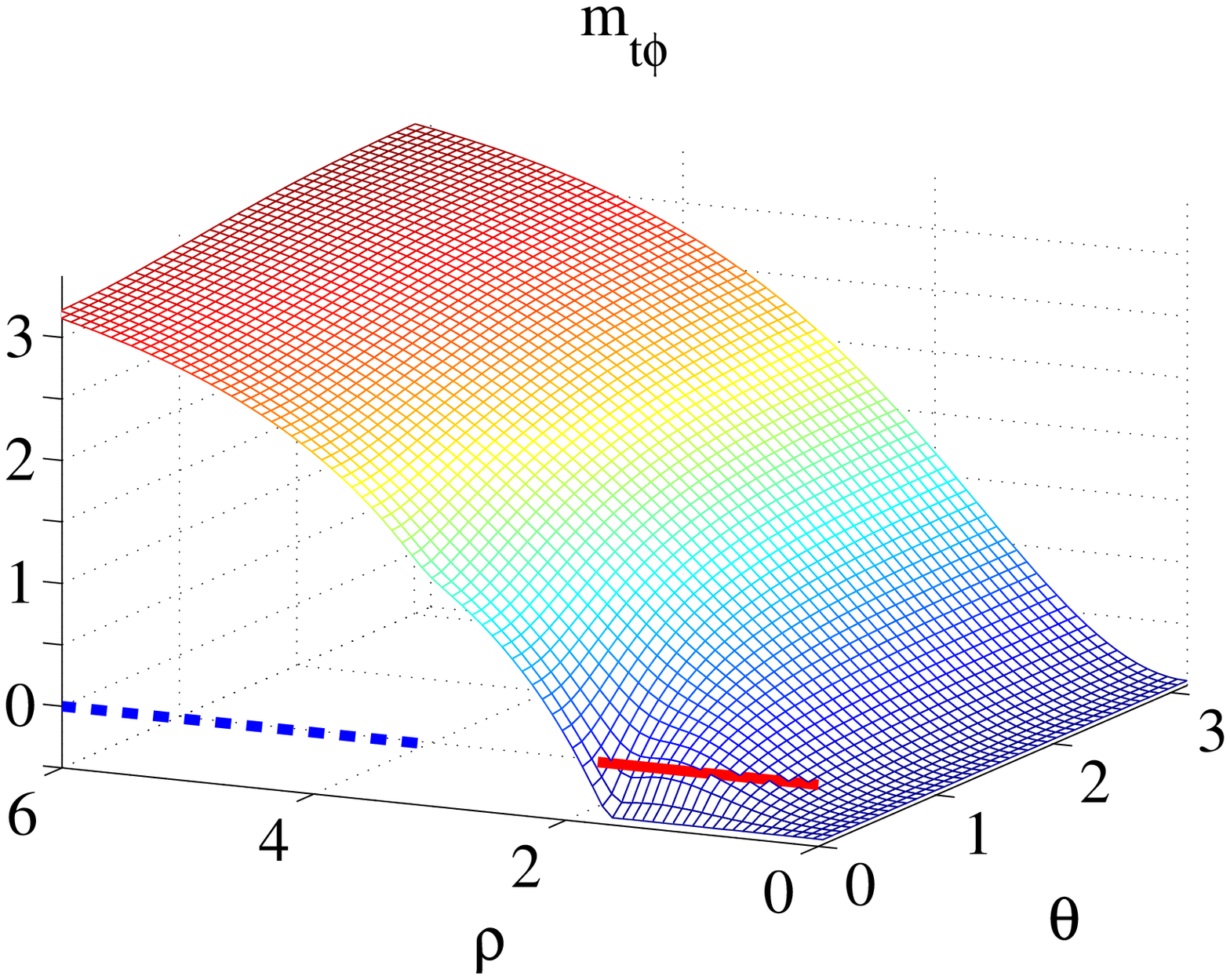}
\hspace{1cm}
\includegraphics[width=7cm]{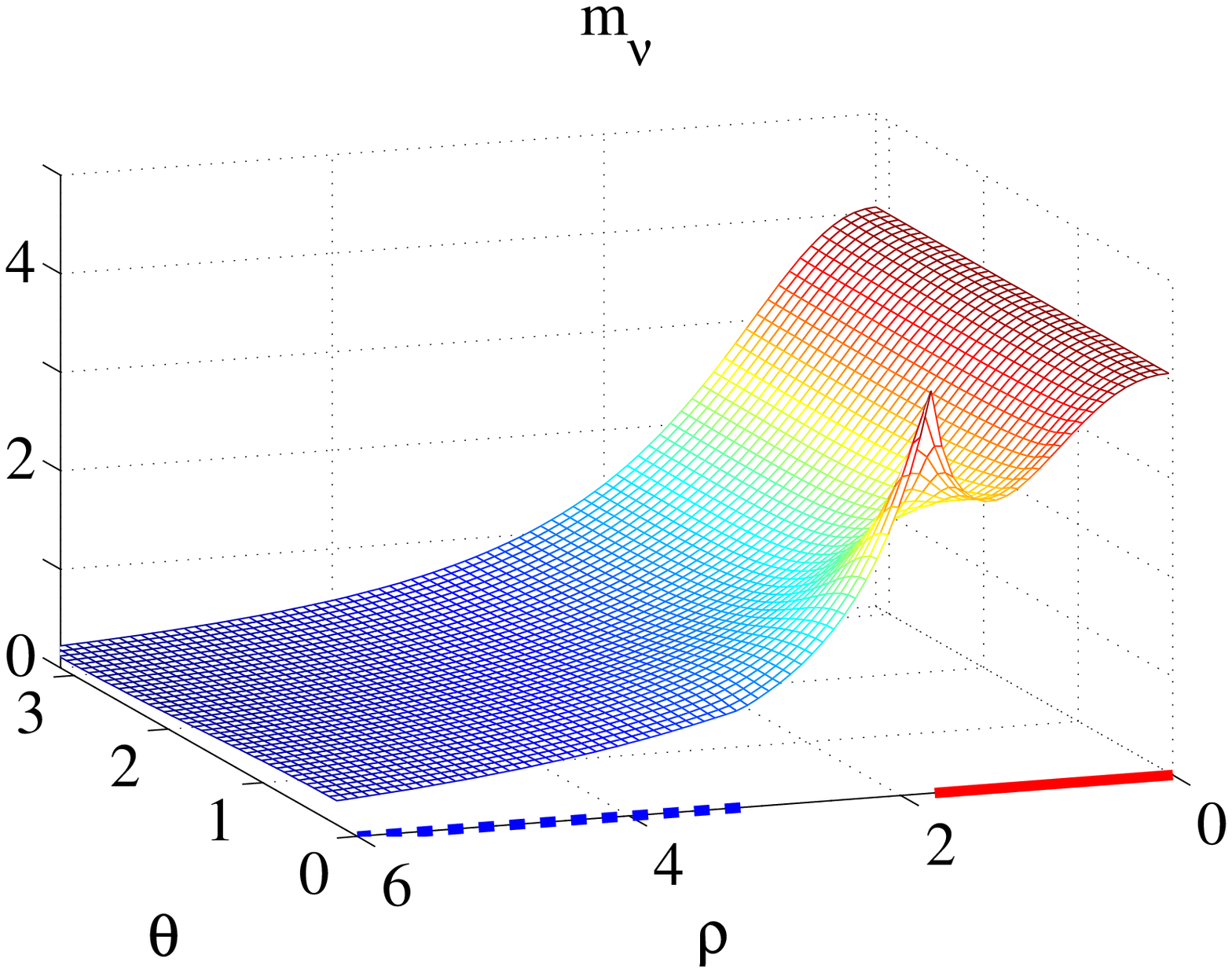}
\end{center}
\caption{
\label{fig:m33etc}
Numerical solutions of $m_{\phi\phi}$, $m_{\psi\psi}$, $m_{t\phi}$ and $m_{\nu}$, corresponding to the solutions at linear order in Fig.~\ref{fig:m13,m23}. 
}
\end{figure}

Based on these considerations, we have solved the second order perturbations numerically. 
An example of solutions is shown in Fig.~\ref{fig:m33etc}. 
The perturbative solutions are constructed with respect to the thin black rings ($c<1/2$) and their physical properties are discussed in the next subsection.  
For the fat black rings ($c>1/2$), the control of the homogeneous modes in Eq. (\ref{eq:m22 -2nd order}) becomes harder, and we could not construct a solution on this branch. 
Thus we restrict ourselves to the doubly spinning thin BRs in the next section. 
From Fig.~\ref{fig:m33etc}, we notice that the solution $m_{\nu}$ has a sharp peak at the boundary between the horizon and inner axis (at $\theta=0$). 
This non-smooth behavior would be inevitable, because the background metric itself is not sufficiently smooth in the canonical coordinates; 
$e^{2\nu_{(bg)}}$ diverges at $z=\pm c \kappa^2, \kappa^2$ on $r=0$.

Before ending this section, we give comments on the accuracy of numerical solutions.
The Smarr relation (\ref{eq:Smarr}) is expanding upto second order and no linear order term appears. As discussed above, the Smarr relation does not suffer from the issue of the constant mode, this relation is useful to check the consistency.   
We have confirmed that the Smarr relation at the second order is well satisfied within $2\%$ relative errors. 
Furthermore, we also found that the relation still holds within the same error even if we do not impose the condition (\ref{eq: m33=m22 condition}). 
This is consistent with the fact that the Smarr relation is satisfied even if the force balance condition $\Delta \phi = \Delta \psi$ is not enforced as discussed above.

\subsection{Perturbative doubly spinning black rings}
\label{Numerical solutions}

By the perturbations upto second order, we have incorporated the backreaction from the second spin. 
The backreaction affects the spacetime geometry, and so we can investigate the physical properties of the doubly spinning black rings based on our solutions for thin black rings.

All physical quantities depend on the periodicity $\epsilon$, so that they are not free from the ambiguity of $\alpha_\nu$ (or the constant mode in $m_\nu$).
To avoid the ambiguity, we adopt the so-called reduced spin variable $j$ and reduced horizon entropy, 
\begin{eqnarray}
j^2 &=&  {\frac{27\pi}{32}} \frac{J_\phi^2 + J_{\psi}^2}{M^3},
\cr
s &=& \frac{3\sqrt{3}}{4\sqrt{\pi}} \frac{S}{M^{3/2}}. 
\end{eqnarray}
Here we have introduced the reduced spin by a sum of squared angular momentum. 
This is useful variable to draw a phase diagram of doubly spinning BRs in a two-dimensional plane, instead of in a three-dimensional space.
$j^2$ does not depend on the periodicity $\epsilon$ by its construction, while the reduced entropy does depend on it.        
However, expanding these quantities upto second order, we find
\begin{eqnarray}
j^2 &=&  
\frac{(c+1)^3}{8 c} 
+ \omega_\psi^2 (
  j^2_{\psi[2]}
+ j^2_{\phi[2]}),
\cr
s &=& 
2 \sqrt{c(1-c)} + \omega_\psi^2 s_{[2]},
\end{eqnarray}
where the second order corrections are 
\begin{eqnarray}
j^2_{\psi[2]}
&\equiv& \frac{   \alpha_{t\psi}^2 c \kappa^2 (1-c)^3 }{32}  , 
 \nonumber
\cr
j^2_{\phi[2]}
&\equiv& 
   \frac{(1+c)^3}{32 c^2 {\kappa}^2} 
   \left(
       8c \kappa^2 \alpha_{t\phi}
     -3(1-c) (\alpha_{\phi\phi}+ \alpha_{\psi\psi})
  \right)  ,
\nonumber
\cr
s_{[2]}
&\equiv&
 \sqrt{c(1-c)} 
\left(   m_{\psi\psi}+ m_{\phi\phi} +m_{\nu}^{\alpha=0}
        -\frac{3(1-c)(\alpha_{\phi\phi} + \alpha_{\psi\psi})}{4c\kappa^2}
       \right), 
\end{eqnarray}
and we also use $(j^2)_{[2]}  = j^2_{\phi[2]}+ j^2_{\psi[2]} $.
$j^2_{\phi[2]}$ and $j^2_{\psi[2]}$ represent contributions from respective angular momenta.
As it shows, the dependence of $\alpha_\nu$ in $s$ is higher order in perturbations so that it is free from the ambiguity at the leading order $O(\omega_\psi^2)$.
For the MP black holes, the reduced entropy is 
\begin{eqnarray}
 s_{MP} = 2 \sqrt{ 1- j_\phi^2 - j_\psi^2 
 + 
 \Bigl(  [1-(j_\phi-j_\psi)^2][1-(j_\phi+j_\psi)^2] \Bigr)^{1/2}
 }, 
\end{eqnarray}
and the horizon exists only when $ j_\psi^2 + j_\phi^2 + 2|j_\phi j_\psi| \le 1$.

\begin{figure}[t]
\begin{center}
\subfigure[]{
\includegraphics[width=8cm]{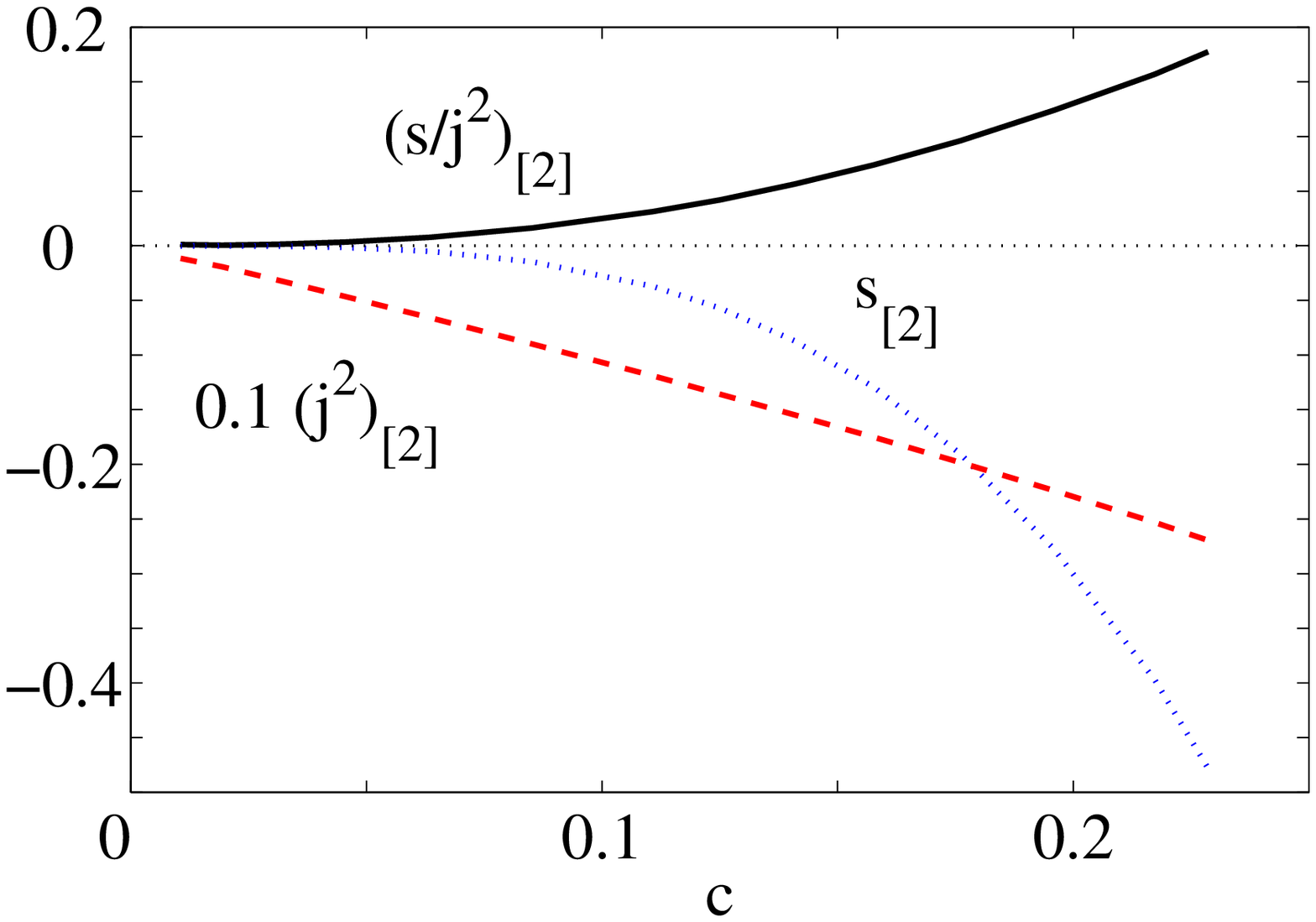}
\label{fig:s2overj2 A}
}
\subfigure[]{
\includegraphics[width=7.5cm]{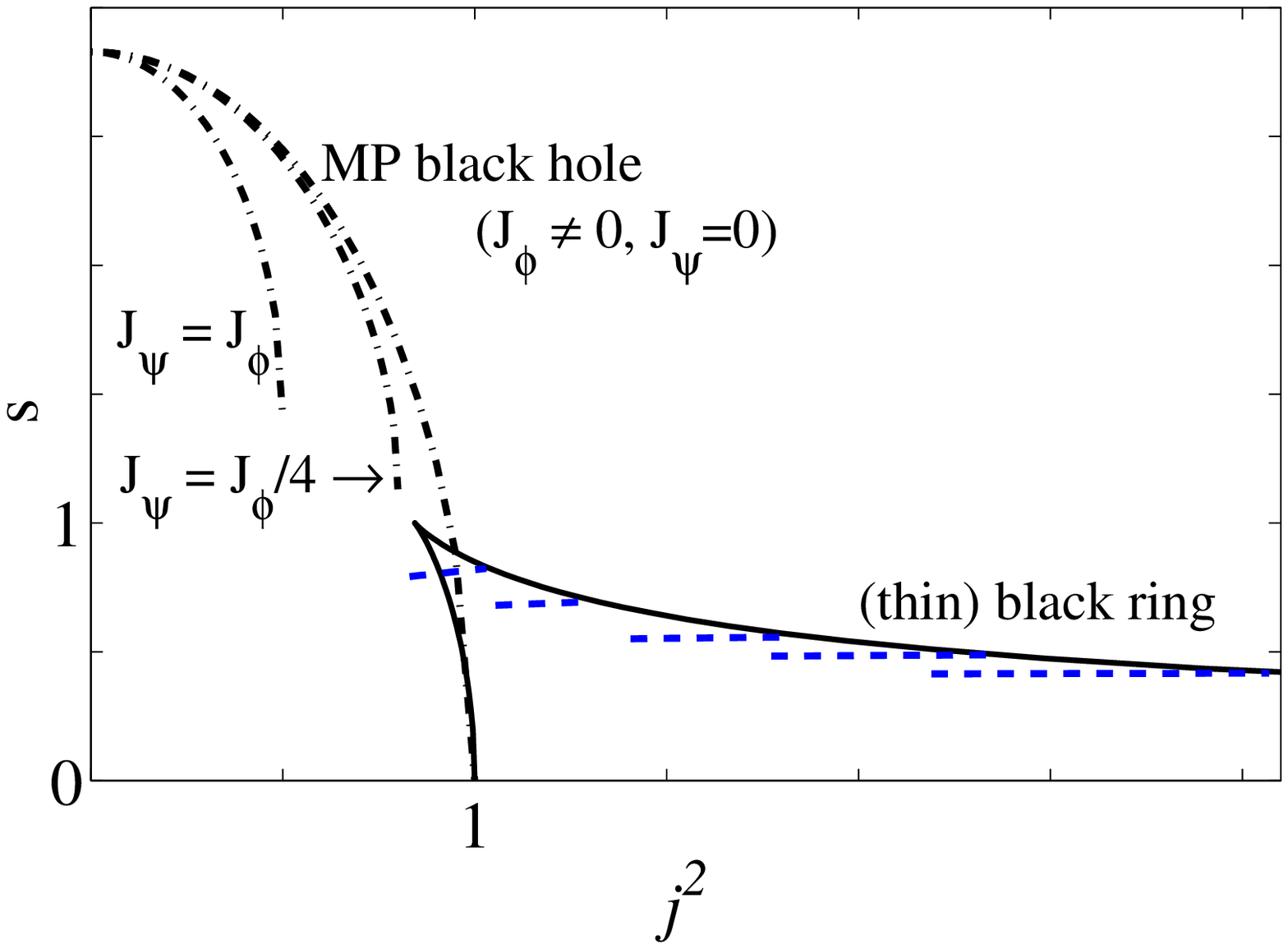}
\label{fig:s2overj2 B}
}
\end{center}
\caption{
\label{fig:s2overj2}
(a) The second order correction in $s$, $s/j^2$ and  $10^{-1}\times j^2$.
(b) Reduced entropy vs $j^2 (=j_\phi^2+j_{\psi}^2)$ for the black rings (solid) and Myers-Perry black holes (dot-dashed). 
The curves for MP black holes correspond to $J_\psi=0$, $J_\psi=J_\psi/4$, and $J_\psi=J_\psi$ cases. Since the existence of horizon requires $j_\phi^2+j_\psi^2 + 2|j_\psi j_\phi| \le 1$, the latter branch ends at some point.
Each dashed lines emanating from the sequence of thin black ring indicates the sequence of doubly spinning black rings. 
}
\end{figure}

In Fig.~\ref{fig:s2overj2}, $s_{[2]}$ and $(j^2)_{[2]}$ are shown. Both of them are negative, implying that the doubly spinning black rings have less entropy for given mass. 
It is a general process to decrease the horizon area as a spin increases, as are the cases in the Kerr and MP black holes.
However, we have no reason to expect smaller reduced angular momentum, and so a question is why $(j^2)_{[2]}$ is negative. 
In the evaluation of $(j^2)_{[2]}$,  $j^2_{\psi[2]}$ gives small positive contribution while  $j^2_{\phi[2]}$ gives dominant negative contributions, i.e. $|j^2_{\phi[2]}|  \gg j^2_{\psi[2]}$.  
Thus the negative value of $(j^2)_{[2]}$ comes from the backreaction to $j_{\phi}^2$ and we have $(j^2)_{[2]} \approx j_{\phi}^2 <0$. 
By comparing with the MP black holes (Fig.~\ref{fig:s2overj2}), the present result might suggest that there is a maximum reduced angular momentum $j^2$ and its maximum value is achieved by the single spinning BRs.  
This does not necessarily mean that the maximum total angular momentum $J^2=J^2_\phi+J^2_\psi$ is saturated by the single spinning black rings.  We expect that the angular momentum of the single spinning black rings provides lower limit of the total angular momentum, because the second angular momentum would have no significant role for the radial balance of forces in such black rings. 
Therefore, the negative reduced angular momentum means that the increase of mass due to the backreaction is more rapid than the increase of spins in these perturbations.

\begin{figure}[t]
\begin{center}
\includegraphics[width=7cm]{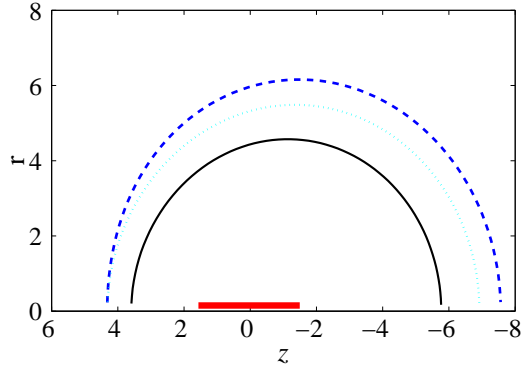}
\end{center}
\caption{
\label{fig:ergoregion}
Ergosurfaces for the background BR with single spin (inner solid curve) and the doubly spinning BR (outer dashed curve with $\omega_\psi=0.3$). 
To compare the two ergosurfaces, we draw the thin dotted line by just rescaling the inner solid curve for the background black ring.
The doubly spinning black rings have larger ergosurface, and the surface tends to spread outside. 
The thick line on $r=0$ represents the horizon. 
}
\end{figure}

Finally, we discuss the ergosurface of the doubly spinning BRs. 
The ergosurface at which $\partial/\partial t$ changes from timelike to spacelike is shown in fig.~\ref{fig:ergoregion} for the background BR and the doubly spinning BRs.  
The doubly spinning black rings have larger ergosurface, and the surface tends to spread outside.  
This behavior is opposite from that observed for a black ring with a rotating 2-sphere \cite{Iguchi:2006tu} which has a conical singularity.

Related to this geometrical property, we give a comment on closed timelike curves. 
The black ring with a rotating 2-sphere \cite{Iguchi:2006tu} has closed timelike curves, i.e. $g_{\psi\psi}<0$, in some parameter space. 
On the other hand, it is clear that closed timelike curves are not allowed as long as we work on the perturbative expansion (\ref{eq:def pert m_ab}), since $g_{\psi\psi}$ and $g_{\phi\phi}$ cannot change the sign by perturbations.
Furthermore, the positive amplitudes of $m_{\phi\phi}$ and $m_{\psi\psi}$ means that the actual perturbations has no symptom of closed timelike curves.

\section{Extensions of the approach}
\label{sec:Extensions of the approach}

So far, we have focused on the perturbations to understand how we solve the boundary value problem (\ref{eq:symbolic eq DDM=S}) for a given rod-structure. 
Besides, by the example of doubly spinning black rings, we have made it clear that the rod-structure can be actually specified by requiring physical properties of an expected black hole.

Clearly, it is possible to extend the perturbative solutions to fully nonlinear solutions by directly solving Eq.~(\ref{eq:symbolic eq DDM=S}). 
Although the fundamental problem of the constant mode in $m_\nu$ remains, all boundary conditions explained in Sec.~\ref{subsec:Asymptotic analysis} and \ref{subsec:Rod structure} can be applied, and an iterative scheme to solve elliptic PDEs are available. 
These techniques are also useful to discuss other interesting systems. 
For instance, several static axisymmetric solutions (black holes or bubbles) will be extended to stationary ones \cite{Emparan:2001wk,Elvang:2004iz}, or as an extremely exotic possibility,  it might be possible to construct a black hole (or black ring) put on the center of the Emparan-Reall's black ring.

Another interesting possibility to apply fully nonlinear numerical method is the search for black rings in higher dimensions. 
There have been no known feasible approaches to find such objects so far. 
Here we discuss a possible approach to construct such object based on numerics. 
A crucial point to construct higher dimensional black rings and other objects is to find an appropriate coordinate system, as is often the case in general relativity.
In particular, it is best in numerics to find rectangular coordinates in which all boundaries coincide with the coordinate lines. 
For higher dimensional black rings, the canonical coordinates (\ref{eq:r2=detg}) is no longer available in $D\ge 6$, but the general conformal form of the metric is still valid.

The metric (\ref{eq:conformal metric anstaz1}) can be written as 
\begin{eqnarray}
  ds^2  
   &=& \sum^{D-n-2}_{\alpha, \beta=1}  g_{\alpha \beta} (r,z) dx^\alpha dx^\beta
       + e^{2\lambda(r,z)} d\sigma_{(n)}^2 
       + e^{2\nu(r,z)} (dr^2 + dz^2).
\label{eq:metric anstaz1}
\end{eqnarray}
Here $d\sigma_{(n)}^2 = \gamma_{ij}dx^i dx^j$ is the metric of $n$-sphere. 
This form of metric will be suited for describing a black object with $S^i \times S^j$ topology. 
In fact, the Minkowski spacetimes for $D\ge 5$ can be embedded in this form:
\begin{eqnarray}
&&
  \sum^{D-n-2}_{\alpha, \beta=1}  g_{\alpha \beta}  dx^\alpha dx^\beta
  = 
  - dt^2 + (\sqrt{r^2+z^2} - z) d\Omega^2_{(D-n-3)},
\nonumber
\\
&&  e^{2\lambda} =  \sqrt{r^2+z^2} + z,
\label{eq:def metric for S^nS^m}
\\
&&  e^{2 \nu } =  \frac{1}{2\sqrt{r^2+z^2}}.  
\nonumber
\end{eqnarray}
The corresponding ring coordinates were first discussed in \cite{Emparan:2006KITP}.
For $D=4$, the metric is the so-called Papapetrou form, 
\begin{eqnarray}
  ds^2 = -dt^2 + r^2 d\phi + dr^2+dz^2. 
\end{eqnarray}
For $D=5$, making the coordinate transformation 
\begin{eqnarray}
  r= \frac{1}{2} {\mathcal R}^2 \sin 2 \Theta ,
\quad
  z= \frac{1}{2} {\mathcal R}^2 \cos 2 \Theta, 
\label{eq:rz-spheroidal def}
\end{eqnarray}
with $0 \le \Theta \le \pi/2$, we get the Minkowski metric in spheroidal coordinates. 
\begin{eqnarray}
 ds^2 = - dt^2 + {\mathcal R}^2 \cos \Theta^2 d\psi
         + d{\mathcal R}^2 + {\mathcal R}^2 (d\Theta^2 + \sin^2\Theta d\phi^2). 
\end{eqnarray}
For $D\ge 6$, the same coordinate transformation as (\ref{eq:rz-spheroidal def}) is available.

Taking these Minkowski metric as a background metric and assuming a general form of metric, like Eq. (\ref{eq:general ansatz for 5D case}), the PDEs are reduced to coupled elliptic PDEs with the Laplace operator $\vec{\nabla}^2$ in three-dimensional cylindrical coordinates. 
The canonical condition (\ref{eq:r2=detg}) is not available anymore, and important properties of rod-structure is also unavailable. 
For instance, as we notice from the above example, the principle of ``one eigenvalue of $g_{ab}$ becomes zero for a given $z$" is broken explicitly on axes, since multiple angular components, say $d\Omega_n^2$, vanishes at the same time. 
However, we can still require physical (or regularity) conditions on the axes and other portion of $r=0$. 
Then it is possible to set up a problem of constructing higher dimensional black rings as a well-defined (numerical) problem. 
(This is analogous to finding a black hole localized on the KK circle in $D=5$, whose construction in numerics had been done without relying on the rod-structure \cite{Emparan:2001wk,Kudoh:2004hs,Kudoh:2003ki}).
Subtle issues will come with rotation of the rings. 
For instance, it will be necessary to factor out angular variables related to sphere in the off-diagonal metric components, and the associated spheres may not be round sphere.
Besides, we may also worry about the ergosurface, at which vanishing of $g_{tt}$ may cause numerical problem.
Nevertheless, higher dimensional BRs are great challenging issues in higher dimensional gravity, and further research is necessary to tackle the issues.

Finally, we end this section with another interesting possible extension of the present approach. From the point of view of numerics, the cosmological constant is not an essential problem once a good coordinate system is provided. 
Here we provide the metric of five-dimensional AdS in the conformal form, 
\begin{eqnarray}
&& ds^2 
= - \left(1 - \frac{\Lambda}{6} h^2 \right) dt^2 
+ \frac{h^2}{\chi^2}
\left[
  \left( \sqrt{r^2+z^2} -z\right) d\phi^2
+ \left( \sqrt{r^2+z^2} +z\right) d\psi^2
+ e^{2\nu} (dr^2+dz^2) 
\right],
\cr
&& h= \chi \left[ 1 + \frac{\Lambda}{24} \chi^2 \right]^{-1}, 
\qquad
  \chi^2 = 2 \sqrt{r^2+z^2}.
\end{eqnarray}
where $e^{2\nu}$ is the same as in (\ref{eq:def metric for S^nS^m})  and $\Lambda<0$ is the cosmological constant. 
The domain of $\chi$ is $\chi = [0, \sqrt{24/|\Lambda|}]$, and so $h=[0,\infty]$.
This metric is found by coordinate transformation from the static coordinates of AdS.

\section{Summary}
\label{sec:Summary and discussion}

We have discussed the formulation of solving stationary axisymmetric spacetimes and focused on the example of doubly spinning BRs. 
The doubly spinning BRs are distinguished from the BRs with single spin by the new independent asymptotic conserved charge.

This example shows that we can formulate the problem in a form suitable for numerical implementation.  
As a first step, we have expanded the nonlinear equations for small second angular velocity and obtained the perturbative equations for the doubly spinning BRs. 
Separation of variables is not available for some set of equations, and we have integrated the perturbative equations numerically. 
The perturbations have been solved upto second order in small angular velocity.
The constructed doubly spinning BRs are regular and free from the conical singularity. 
However, the absolute value of periodicity of axes, $\Delta \phi = \Delta \psi=2\pi \epsilon$, cannot be determined for each solution in the present approach.  
The ambiguity comes from the constant mode in $e^{2\nu}$.

Although we have discussed the perturbations in the canonical coordinates, it is not necessary to work in these coordinates if we are interested only in the perturbative treatment of doubly spinning BRs. The corresponding analysis in the ring-coordinates will be more tractable than in the canonical coordinates.

To discuss the physical properties of doubly spinning BRs, we have evaluated the reduced entropy and reduced angular momentum for each perturbative solution.
As peculiar to rotating black holes, the second spin associated to the $S^2$-sphere deceases the entropy for a fixed mass and angular momenta. 
On the other hand, the sum of reduced angular momenta decreases. 
Our construction of doubly spinning BRs is only for the branch of thin black rings, and it is unclear how the fat black rings behave with the second spin. 
An interesting question is whether the region where the Myers-Perry (MP) black holes and black rings overlap disappears or not. 
Presumably, the overlapped region will disappear for some range of second spin, because the branch of doubly spinning MP black holes with regular horizon leaves the overlapped region and its reduced entropy is bounded below (see Fig.~\ref{fig:s2overj2}). 
If the doubly spinning BRs have smaller reduced entropy all the way down to its (expected) maximum spin, they can never reach the parameter region of MP black hole in some range of $j^2$ (or in three-dimensional space of $\{ j_1^2, j_2^2, s\}$).

The issue of dynamical stability of black rings and MP black holes are also important. 
The thin black rings will be unstable via Gregory-Laflamme instability along the $S^1$ direction~\cite{Emparan:2001wn,Hovdebo:2006jy}, and the fat black rings have been though of more unstable~\cite{Arcioni:2004ww,Nozawa:2005eu}. 
Furthermore, the doubly spinning black rings presumably suffer from superradiant instabilities peculiar to a black hole with a rotating sphere \cite{Dias:2006zv}.

We have also discussed interesting possible applications of numerics to constructing higher dimensional black rings and black rings in AdS. 
Although there are some possible technical subtleties involved in the approach, the direct approach to constructing new black rings is the great challenging. 
All the metric discussed above will be useful for our future study of black rings in higher dimensions and/or with the cosmological constant.
We will be back these open problems in our future publications.

\section*{Acknowledgements} 
The author would like to thank to G.~Horowitz, A. Virmani,
T. Mishima, S. Tomizawa and D. Mateos for useful discussion and comments. 
He is also grateful to the KITP and the workshop ``Scanning New Horizons: GR Beyond 4 Dimensions'' in Santa Barbara for the stimulating environment. 
HK is supported by the JSPS (No.~10059). 

\bibliographystyle{unsrt} 


\end{document}